\documentclass[10pt]{article}

\newcommand{\diag}{\mbox{\sf diag}}

\newcommand{
\cnt}[1]{
\begin{center}#1
\end{center}}

\newcommand{\spc}{
\hspace{5mm}}

\newcommand{\et}{
\hspace{5mm}\mbox{and}
\hspace{5mm}}

\newcommand{\ou}{
\hspace{5mm}\mbox{where}
\hspace{5mm}}

\newcommand{\avec}{
\hspace{5mm}\mbox{with}
\hspace{5mm}}

\newcommand{\armn}[2]{\widehat{#1}=\mbox{\sf argmin}_{#1} \; \left\{
{#2}\right\}}

\newcommand{\dm}[2]{\mbox{\footnotesize $ \, [{#1} \times {#2}]
$}}

\newcommand{\rf}[1] {{\bf(\protect\ref{#1})}}

\newcommand{
\beq} [1] {
\begin {equation} \label {#1}}

\newcommand {\eeq
}{
\end{equation}}

\newcommand{\btb}[3] {\protect
\begin{table}[hbtp]
\begin{center} \protect\caption{#1} \label{#2}
\vspace{1ex} \small
\begin{tabular} {#3}}

\newcommand{\etb}{
\end{tabular}
\end{center}
\end{table}}

\newcommand{\upb}[2]{
\begin{array}c{#1} \\
{#2}
\end{array}}

\newcommand{\upa}[1]{\upb{#1}{}}

\newcommand{\upc}[3]{\upb{#1}{\dm{#2}{#3}}}

\newcommand{\myddate}{Sep. 6, 2007 @06:44}

\newcommand{\eff}{\mbox{\sf Efficacy}}

\newcommand{\ccv}{\mbox{\sf Cov}}

\newcommand{\vov}{\mbox{\sf DiaCov}}

\newcommand{\sdm}[2]{,\spc \mbox{\small Dim.}: \dm #1 #2}

\newcommand{\bp}{\mbox{\sf BP}}

\newcommand{\kbp}{k_{\footnotesize \bp}}

\newcommand{\rh}[1]{Eq. \rf{#1}}

\newcommand{\rhs}[1]{Eqs. \rf{#1}}

\newcommand{\frrac}[2]{#1/#2}

\newcommand{\sub}[2]{{#1}_{\mbox{\footnotesize \sf #2}}}

\newcommand{\sth}[1]{\spc \mbox{\sf such that} \left\{#1\right\}}

\newcommand{\lgd}[2]{\mbox{\footnotesize
\begin{tabular}{|c|}
\hline
#1 \\
Rank #2 \\
\hline
\end{tabular}}}

\newcommand{
\skp}[1] {}

\newcommand{\skmath}[1] {}

\newcommand{
\tmpry}[1] {}

\begin{document}

\cnt{{\Large \bf Total singular value decomposition. \\
Robust SVD, regression and location-scale} \\
William J.J. Rey\footnote{Author's address: Fontaine des Gattes
31, 1470 Bousval, Belgium. Email: w\_rey@tvcablenet.be \\
Dr. Rey is retired from `Philips Research Laboratories' and `Eurandom',
Eindhoven, The Netherlands.

Last edited: \myddate. \\
This is a revised version of the text dated Jun. 1, 2007. Besides
minor text corrections, the notation has been changed to be more
in agreement with published material.

}}

\begin{abstract}

Singular Value Decomposition (SVD) is the basic body of many statistical
algorithms and few users question whether SVD is properly handling
its job.

SVD aims at evaluating the decomposition that best approximates
a data matrix, given some rank restriction. However often we are
interested in the {\em best components} of the decomposition rather
than in the {\em best approximation}. This conflict of objectives
leads us to introduce {\em Total SVD}, where the word ``Total"
is taken as in ``Total" least squares.

SVD is a least squares method and, therefore, is very sensitive
to gross errors in the data matrix. We make SVD robust by imposing
a weight to each of the matrix entries. Breakdown properties are
excellent.

Algorithmic aspects are handled; they rely on high dimension fixed
point computations.

\end{abstract}

\section {Introduction}

The presented approach goes upside-down. Indeed, starting with
our ultimate goal, SVD, we precisely define the ingredients we
need and, then, we keep them in mind in the course of actions.
The two main principles will be

\begin{itemize}

\item {\em Be robust}. This will be understood as limiting the
effect of each observation on the estimations in a way such that
it cannot unduly pull on the result.

\item {\em Keep it simple}. We would appreciate to plunge the
Singular Value Decomposition (SVD) problem in a M-estimation setup.
However, this does not seem possible without imposing heavy restrictions
on the data matrices. We will remain pragmatic and be oriented
toward the numerical aspects.

\end{itemize}

We observe that our natural use of weights is fully consistent
with the theories developed in the field of robustness. Without
any effort, we come with the concepts of Breakdown point and M-estimation.

\subsection {SVD}

There are many ways to look at data matrices and, thinking at
principal components, most of the ways favor either the rows or
the columns of the data matrix; they are either observation- or
variable-oriented. Some analysis methods try to be more balanced,
for instance the (non-robust) biplot as defended by Bradu and
Gabriel (1978), Gabriel (1998), Gower (2004) or Le Roux and Gardner
(2005). Considering a $ \dm m n$-data matrix $X$, a low-rank approximation
is estimated (typically of rank 2) and graphically presented.
Strangely, the approximation is worked out without taking into
account the scatter of the matrix entries. Let us explain.

The singular value decomposition
\[ \upc X m n \upa {\approx} \upc U m p \upc {\Lambda} p p \upc
{V',} p n \spc \upa {p \leq \min \{m, n \}} \]
where
\[ U' \; U = V' \; V = I_p \et \Lambda = \diag \{\lambda_1, ...,
\lambda_p \}, \; \lambda_1 \geq ... \geq \lambda_p \geq 0 \]
is conveniently worked out with the help of an intermediate step
\[ X \; {\approx} \; A \; {B'} = U \; {\Lambda} \; {V'} \]
and an alternating computational scheme. This paper is centered
on the approximation
\beq {eq1} \upc X m n \upa {\approx} \upc A m p \upc {B'} p n
\eeq
and essentially ignores the last step,
\[ \upc A m p \upc {B'} p n \upa = \upc U m p \upc {\Lambda} p
p \upc {V',} p n \]
seeing that it does not involve any approximation and can be performed
by any (least squares) method. Works on alternating computational
schemes are referred to in Ke and Kanade (2005) and Croux et al.
(2003), for instance, and have flourished since the criss-crossing
of Bradu and Gabriel (1978).

The goal is to estimate $A \; B'$ as it appears in \rh{eq1} by
a minimisation procedure
\beq {eq2} \armn {A \; B'} {\parallel X - A \; B' \parallel} \eeq
where $ \parallel . \parallel $ is a norm. This problem is solved
by alternating between
\beq {eq3} A \; \mbox {being known}, \spc \armn {B'} {\parallel
X - A \; B' \parallel} \eeq
and
\beq {eq4} B \; \mbox {being known}, \spc \armn A {\parallel X
- A \; B' \parallel} . \eeq
Unfortunately, the norm in \rh{eq2} is ill-defined and, except
in the least squares set-up, it has little to do with those of
\rh{eq3} and \rh{eq4}. Moreover, the two latter are inconsistent
seeing that, at the end of the convergence, the two errors with
respect to the approximated entry $X_{i j} $, namely
\[ [X_{i j} - (A \; \hat B')_{i j}] \; \mbox {given $A$} \et [X_{i
j} - (\hat A \; B')_{i j}] \; \mbox {given $B$} \]
often differ. In fact, when robust estimation is of concern, very
generally, they differ.

\subsection{Regression}

Besides the inconsistency of norms, an other little observed hindrance
appears in the solving of \rh{eq3} and \rh{eq4}. For the former,
we wrote {\em $A$ being known} and, of course, it is {\em estimated}
rather than {\em known}. Considering the $ n$ columns of $X$ in
\rh{eq3}, the $ n$ columns of $B'$ are solution of
\[ A \; \mbox {being estimated}, \spc \armn {(B')_{.j}} {\parallel
x_j - A \; (B')_{.j} \parallel} \]
where $ X = (x_1, ..., x_n) $. In the more familiar regression
notation, this is
\[ D \; \mbox {being estimated}, \spc \armn {\beta} {\parallel
y - D \; \beta \parallel} \]
under an error-in-variables model.

\mbox{}

As seen by Huber (1973), the robust treatment of linear regression
is a simple extension of his approach in (1972, 1996) in the location-scale
set-up.

\subsection{The location-scale set-up}

This is at the root of all investigations in robustness and, in
Section 2, we report a very simple-minded view that is exceptionally
rich in spite of being little traditional. In some sense, we address
the reader as if he was not already familiar with all developments
made in the robustness field.

The location-scale set-up is further extended to handle regressions
and this let unexpected features appear in Section 3. Section
4 concludes the exposition of the approach with the SVD treatment.

\section{The location-scale set-up}

\subsection{The weighted approach.}

We consider a set of $ m$ observations $ \{x_1, ..., x_m \} $
that are, as we all assume, distributed according to a so-called
Gaussian distribution. We are interested in their location and
their scatter, entities that can be assessed by their mean $ n$
and their standard deviation (again, `as we all know'), namely
by
\beq {eq5} n = \frac 1 m \; \sum_{i = 1}^ m x_i \et s_x = \left
(\frac 1 m \; \sum_{i = 1}^ m (x_i - n)^2 \right)^ {1 / 2}
. \eeq

Some might prefer to divide by $ (m - 1) $ rather than by $ m$
in the second expression but, for the time being, we consider
this idea as a sheer refinement. Rather, we are concerned by the
possibility of gross errors of estimation due to possible errors
in our data set $ \{x_1, ..., x_m \} $. This is what robustness
is about. To avoid that a few observations pull to much the mean
toward them, it suffices to limit their influences on, for instance,
the sum $ \sum_{i = 1}^ m x_i $ in \rh{eq5}. Inserting weights
$w_i$, the sum becomes $ \sum_{i = 1}^ m w_i \; x_i $.

Clearly, the weighting must have little effect on the `central'
observations and must bound the extreme observations. Assuming
that we bound to some constant $c$, this gives
\beq {eq6} w (x) = \left\{
\begin {array} {ll} \frac c {n - x}, & \mbox {if $x\ll m$}, \\
1, & \mbox {if $|x-m|\ll s$}, \\
\frac c {x - n}, & \mbox {if $x\gg m$} .
\end {array} \right.
\skp {abc \} .} \eeq
Remark, already at this level, that the weights are allocated
with help of the mean and standard deviation estimates. Further
on, we apply the weight definitions
\beq {eq7} w (x) = \left(1 + | u |^q \right)^ {- 1 / q} \ou
u = \frac {x - n} {k_1 \; s_x}, \spc 0 < q < \infty \eeq
and, at the limit $q = \infty$,
\[ w (x) = \left\{
\begin {array} {ll} 1, & \mbox {if $|u|\leq 1$}, \\
\frrac 1 {| u |}, & \mbox {if $|u|\geq 1$} .
\end {array} \right.
\skp {abc \} .} \ou u = \frac {x - n} {k_1 \; s_x}, \]
that imply for the above constant
\[ c = \lim_{| x - n | \rightarrow \infty} \; w (x) \; | x | =
k_1 \; s_x. \]
The so-called `tuning constant' $k_1$ controls the level of robustness
that is desired. The parameter $q$ will be numerically tried at
five different values, $ 1, 2, 4, 8$ and $ \infty$, in order to
select the most appropriate one. The weights of $q = 1$ were already
implemented in the numerical package of Klema (1978) and they
are presented by Coleman et al. (1980) under the name `Fair'.
At $q = \infty$, the weighting is identical to what Huber (1964)
found as optimal in the near-vicinity of the Gaussian distribution.

Clearly having inserted weights, \rh{eq5} must be adapted and,
at first sight, the next writings could appear appropriate
\[ n = \frac 1 {\sum_{i = 1}^ m w_i} \; \sum_{i = 1}^ m w_i \;
x_i \et s_x^2 = \frac 1 {\sum_{i = 1}^ m w_i^2} \; \sum_{i =
1}^ m [w_i \; (x_i - n)]^2 . \]
The above estimator of the population variance is severely biased
on two grounds: On the one hand, it is greatly underestimating
$s_x^2$ due to systematically down-weighting the large terms $
(x - n)^2$ and, on the other hand, we have omitted to take the
correct number of degrees freedom into account. We now deal with
these two issues.

The severe under-estimation is taken into account by inserting
a correction factor
\beq {eq8} k_2^2 = \frac {\int_{- \infty}^ {\infty} w (x)^2
\; m (x) d \, x} {\int_{- \infty}^ {\infty} w (x)^2 \; x^2 \;
m (x) \; d \, x} \ou m (x) = \exp^{- x^2 / 2} / \sqrt {2 \; \pi}
. \eeq
Rather than dividing by $ m$, at least in the linear context,
we know that we must take into account a reduced number of degrees
of freedom. In this approach with weights, it is less clear how
the situation must be handled. We have opted for a correction
term $ \frac N {N - 1} $ where $N$ stands for the `effective'
number of observations, the number of observations that play a
role in the variance estimate. Very generally, whatever an item
$ (u_i) $ can be, a weighted average takes the form $ \sum_{i
= 1}^ m w_i \; (u_i) / {\sum_{i = 1}^ m w_i} $, therefrom come
the expressions of the average weight $ \bar w$ and the `effective'
sample size $N$
\[ \bar w = \frac 1 {\sum_{i = 1}^ m w_i} \; \sum_{i = 1}^ m
w_i \; (w_i) \et N = \frac {\sum_{i = 1}^ m w_i} {\bar w} = \frac
{(\sum_{i = 1}^ m w_i)^2} {\sum_{i = 1}^ m w_i^2} . \]
Further on, we compare estimators with respect to how many observations
influence their taken values. Very simply, this can be measured
by
\beq {eq9} \eff = \frac N m. \eeq

It is now time to gather the various items we met with. The mean
$ n$ and the scatter $s_x$ are the solution of a fixed-point problem
where, given previous estimates $ (n, s_x) $, the weights and
the new $ (n, s_x) $ are estimated according to
\beq {eq10} \left\{
\begin {array} l \mbox {Given} \; n \; \mbox {and} \; s_x, \\
\mbox {} \\
w_i = \left(1 + | u_i |^q \right)^ {- 1 / q}, \ou \; u_i =
\frrac {(x_i - n)} {(k_1 \; s_x)}, \\
\mbox {} \\
n = \frac 1 {\sum_{i = 1}^ m w_i} \; \sum_{i = 1}^ m w_i \;
x_i, \\
\mbox {} \\
s_x^2 = k_2^2 \; \frac 1 {\sum_{i = 1}^ m w_i^2} \; \sum_{i =
1}^ m w_i^2 \; (x_i - n)^2 .
\end {array} \right.
\skp {abc \} .} \eeq
The unbiased estimator of the variance is
\[ \widehat {\sigma_x^2} = \frac N {N - 1} \; s_x^2 \ou N = \frrac
{\left(\sum_{i = 1}^ m w_i \right)^2} {\sum_{i = 1}^ m w_i^2}
. \]
For $k_1 \rightarrow \infty$, the above expressions collapse in
the familiar least squares formulae
\[ \left\{
\begin {array} l \mbox {Whatever a priori values} \; n \; \mbox
{and} \; s_x, \spc w_i = 1 \\
\mbox {} \\
n = \frac 1 m \; \sum_{i = 1}^ m x_i \\
\mbox {} \\
s_x^2 = \frac 1 m \; \sum_{i = 1}^ m (x_i - n)^2, \ou k_2 =
1.
\end {array} \right.
\skp {abc \} .} \]
We observe that the set of equations \rf{eq10} in terms of the
two variables $ n$ and $s_x$ describes a contracting mapping when
the weights satisfy \rh{eq6}. Hence, the numerical convergence
can easily be accelerated in the vicinity of the fixed-point solution.

\subsection{Theory and numerical experiments}

Some aspects of the theory will be briefly sketched and further
details can be found in very many places as, for instance, in
Rey (1978, 1983). They will often be supported by numerical experiments.

\subsubsection{Contamination model.}

Rather than assuming that the variates $x_i$ are distributed according
to a nominal probability density function $ \sub f {main} $, Tukey
(1960) considers that a few observations could possibly be drawn
from an other distribution $ \sub f {rubbish} $,
\[ x_i \propto f, \spc f = (1 - \epsilon) \; \sub f {main} + \epsilon
\; \sub f {rubbish} \ou \epsilon \ll \frac 1 2. \]
The contaminated distribution is a mixture distribution. The objective
of robust estimation consists in being as little sensitive as
possible to $ \sub f {rubbish} $. As is usual, we assume a Gaussian
distribution and outliers
\[ \sub f {main} = {\cal N} (\mu, \sigma^2) \et \sub f {rubbish}
\ll \gg {\cal N} (\mu, \sigma^2) . \]
\[ \]
\subsubsection{M-estimation.}

The above weights were introduced as an intuitive manner of limiting
the influence of any extreme observation on the estimates of the
mean and the variance. They also appear very naturally in the
context of M-estimation.

Many estimators can be seen as solutions of minimization problems
and, in the context of location estimation, it takes the form
\beq {eq11} \armn {\theta} {\sum_{i = 1}^ m \rho (x_i - \theta)}
. \eeq
Then, under differentiability conditions,
\[ \sum_{i = 1}^ m \Psi (x_i - \theta) = 0 \ou \Psi (u) = \frac
d {d \, u} \rho (u) . \]
In order to let the weights appear, we transform the left hand
member,
\[ \sum_{i = 1}^ m \Psi (x_i - \theta) = \sum_{i = 1}^ m \; \left
[\frac {\Psi (x_i - \theta)} {x_i - \theta} \right] \; (x_i -
\theta) = \sum_{i = 1}^ m w (x_i) \; (x_i - \theta) = 0 \]
and see that the weights
\beq {eq12} w (x_i) = \frac {\Psi (x_i - \theta)} {x_i - \theta}
\eeq
naturally appears in a minimization framework.

In order to have a minimum, \rh{eq11} must be convex and this
is certainly realized when each of its terms in $ \rho (.) $ is
convex. Then, seeing \rh{eq12}, the restriction \rf{eq6} on the
weights turns out.

All least squares approaches are based on $ \rho (u) = u^2$ and
that corresponds with no weighting (or equal non-zero weights).

\subsubsection{Breakdown point.}

This is a concept developed by Hampel (1968) that has a natural
motivation. Robust estimators remain stable in spite of (very)
extreme observations in the sample. The breakdown is a measure
of how many very extreme observations can be tolerated.

To illustrate, consider the weighted mean estimator
\[ n = \frac 1 {\sum_{i = 1}^ m w_i} \; \sum_{i = 1}^ m w_i \;
x_i \]
and suppose that you add $k$ very extreme points on the right
side of $ n$. Seeing \rh{eq6}, the estimator becomes
\[ n (m, k) = n \approx \frac 1 {\sum_{i = 1}^ m w_i} \; \left
[\sum_{i = 1}^ m w_i \; x_i + k \; c \right], \]
the approximation being valid as long as the point addition little
influences the values of the weights $w_i$. The breakdown point,
\bp, is the fraction such that $ \mbox {\sf Offset}_n $ be very
large or, precisely
\beq {eq13} \bp_a = \lim_{m \rightarrow \infty} \frac {\kbp} {m
+ \kbp} \sth {\mbox {\sf Offset}_n = a} . \eeq
The exact value of $a$ is immaterial as indicated in Note \ref{not2}.

\subsubsection{Numerical convergence}

How fast the iterative process \rf{eq10} converges is of great
practical importance. Noting by $l$ the iteration number, we observe
that the mapping
\[ ... \rightarrow (n^{l - 1}, s_x^{l - 1}) \rightarrow (n^l,
s_x^l) \rightarrow (n^{l + 1}, s_x^{l + 1}) \rightarrow ...
\rightarrow (n^{\infty}, s_x^{\infty}) \]
is contracting. The rates $b_n$ and $b_s$ indicate the convergence
speed,
\beq {eq14} n^{l + 1} - n^{\infty} \approx b_n \; (n^l - n^
{\infty}) \et s_x^{l + 1} - s_x^{\infty} \approx b_s \; (s_x^l
- s_x^{\infty}), \eeq
when process \rf{eq10} is not accelerated. In the near-vicinity
of $ (n^{\infty}, s_x^{\infty}) $ and for symmetrical distributions,
expressions \rf{eq14} are exact.

\subsubsection{Numerical experiments}

All the reported experiments are with respect to the standard
Gaussian distribution and this is further documented in Note \ref{not1}.
The tabulated results are asymptotic, $ m \rightarrow \infty$,
as described in Note \ref{not3}.

It is time to focus on better specifying the weight function given
by the family \rf{eq7}. Clearly, any given weight definition has
consequences on the number of outliers that can be tolerated without
dramatically spoiling the estimations. Hence, the breakdown point
appears to be a natural gauge. Of course, comparing the performances
of various powers must be done under similar conditions. Table
\ref{t1} proposes a summary of the observations and the power-value
$q = 4$ turns out to be an interesting selection, although results
with $q = 2$ or $q = 8$ are little different.

\btb {Selecting the power $q$ in the weight definition \rf{eq7}.}
{t1} {|ccc|}
\hline
\eff & $q$ & $ \bp_1$ \\
\rh{eq9} & \rh{eq7} & \rh{eq13} \\
\hline
0.80 & 1 & 0.276 \\
& 2 & 0.352 \\
& 4 & 0.375 \\
& 8 & 0.382 \\
& $ \infty$ & 0.383 \\
0.90 & 1 & 0.195 \\
& 2 & 0.294 \\
& 4 & 0.329 \\
& 8 & 0.339 \\
& $ \infty$ & 0.342 \\
0.95 & 1 & 0.124 \\
& 2 & 0.235 \\
& 4 & 0.282 \\
& 8 & 0.293 \\
& $ \infty$ & 0.297 \\
\hline
\etb

Table \ref{t2} reports further details with the selected weight
definition,
\beq {eq15} w (x) = \left(1 + u^4 \right)^ {- 1 / 4} \ou u =
\frac {x - n} {k_1 \; s_x} . \eeq
The parameters $k_1$ and $k_2$ control the degree of robustness
of estimation by \rh{eq10}. As indicated by the last two columns
of Table \ref{t2}, the convergence of process \rf{eq10} is somewhat
slow and acceleration is welcome. We observe at the last rows
that a small loss of Efficacy already yields a serious protection
against possible outliers.

\btb {Data with the weights of \rh{eq15}} {t2} {|l|ccc|c|cc|}
\hline
\eff & $k_1$ & $ k_2$ & $k_3$ & $ \bp_1$ & $ b_n $ & $b_s $ \\
\rh{eq9} & \rh{eq7} & \rh{eq8} & $k_1 \, \times \, k_2$ & \rh{eq13}
& \rh{eq14} & \rh{eq14} \\
\hline
0.5 & 0.0987 & 3.7227 & 0.3673 & 0.4110 & 0.7220 & 0.4777 \\
0.6 & 0.1497 & 3.0541 & 0.4571 & 0.4062 & 0.6810 & 0.4723 \\
0.7 & 0.2244 & 2.5258 & 0.5669 & 0.3960 & 0.6275 & 0.4648 \\
0.75 & 0.2760 & 2.2954 & 0.6336 & 0.3881 & 0.5933 & 0.4592 \\
0.80 & 0.3428 & 2.0798 & 0.7130 & 0.3754 & 0.5515 & 0.4513 \\
0.85 & 0.4336 & 1.8730 & 0.8122 & 0.3575 & 0.4984 & 0.4391 \\
0.90 & 0.5686 & 1.6673 & 0.9480 & 0.3294 & 0.4265 & 0.4178 \\
0.92 & 0.6456 & 1.5825 & 1.0217 & 0.3138 & 0.3893 & 0.4043 \\
0.94 & 0.7472 & 1.4937 & 1.1161 & 0.2936 & 0.3442 & 0.3853 \\
0.96 & 0.8942 & 1.3973 & 1.2494 & 0.2662 & 0.2869 & 0.3563 \\
0.98 & 1.1537 & 1.2843 & 1.4817 & 0.2246 & 0.2063 & 0.3036 \\
0.99 & 1.4227 & 1.2116 & 1.7238 & 0.1885 & 0.1458 & 0.2510 \\
1.00 & $ \infty$ & 1.0000 & $ \infty$ & 0.0000 & 0.0000 & 0.0000
\\
[6pt] 0.6418 & 0.1773 & 2.8198 & 0.50 & 0.4028 & 0.6605 & 0.4695
\\
0.8202 & 0.3758 & 1.9955 & 0.75 & 0.3690 & 0.5317 & 0.4471 \\
0.9145 & 0.6227 & 1.6059 & 1.00 & 0.3184 & 0.4001 & 0.4084 \\
0.9601 & 0.8948 & 1.3970 & 1.25 & 0.2660 & 0.2867 & 0.3562 \\
0.9811 & 1.1742 & 1.2774 & 1.50 & 0.2216 & 0.2009 & 0.2994 \\
0.9907 & 1.4516 & 1.2056 & 1.75 & 0.1850 & 0.1405 & 0.2456 \\
0.9953 & 1.7234 & 1.1605 & 2.00 & 0.1559 & 0.0991 & 0.1985 \\
\hline
\etb

\section {Regression} As indicated in the introduction, we consider
the regression model
\[ \upc y n 1 \upa = \upc D n p \upc {\beta} p 1 \upa + \upc {\mbox
{error}} n 1 \]
where the solution $ \hat \beta$ satisfies
\[ D \; \mbox {being estimated}, \spc \armn {\beta} {\parallel
y - D \; \beta \parallel} \]
under an error-in-variables model. It is a topic that includes
quite a few refinements that we here ignore inasmuch as feasible
(by implicitely assuming independency properties).

\subsection{Generalised least squares.}\label {lgls}

The ordinary least squares approach is based on the norm
\[ \parallel y - D \; \beta \parallel^2 = (y - D \; \beta) ' \;
(y - D \; \beta) = \sum_{i = 1}^ n (y_i - d_i' \; \beta)^2 \avec
D = \left(
\begin {array} c d_1' \\
... \\
d_n' \\
\end {array} \right), \]
a norm where the row-vectors $d_i'$ are seen as fully known. The
terms $ (y_i - d_i' \; \beta)^2$ measure the fit-errors and,
when we desire to take into account the total error, we must add
the errors due to the variability of the row-vectors $d_i'$. Namely,
we must complete the terms into
\[ (y_i - d_i' \; \beta)^2 + \beta' \; S_i \; \beta \ou S_i =
\ccv (d_i') . \]

Thinking at robustness, in the weighted case the form
\[ \sum_{i = 1}^ n (y_i - d_i' \; \beta)^2 \spc \mbox {becomes}
\spc \sum_{i = 1}^ n w_i^2 \; (y_i - d_i' \; \beta)^2 + \beta'
\; \left(\sum_{i = 1}^ n w_i^2 \; S_i \right) \; \beta \]
and the above norm $ \parallel y - D \; \beta \parallel^2 $ definition
eventually is modified into
\[ (y - D \; \beta) ' \; W \; (y - D \; \beta) + \beta' \; S_D
\; \beta, \ou S_D = \sum_{i = 1}^ n w_i^2 \; S_i. \]
Note that matrix $W$ is with respect to the squared weights,
\[ W = \diag \{w_1^2, ..., w_n^2 \} . \]

Thus, the generalised least squares approach takes the form
\[ \armn {\beta} {(y - D \; \beta) ' \; W \; (y - D \; \beta) +
\beta' \; S_D \; \beta} \]
\skmath{Normal equation
\[ - D' \; W \; (y - D \; \beta) + S_D \; \beta \]
and
\[ D' \; W \; y = D' \; W \; D \; \beta + S_D \; \beta \]
}that yields to the estimator
\[ \hat {\beta} = J^{- 1} \; D' \; W \; y \]
\skmath{

Evaluation of the covariance
\[ (4.36) : \ccv (\hat \beta) = \frac N {N - 1} \; \sum_{i = 1}
^n u_i^2 \left(\frac {\partial} {\partial u_i} \hat {\beta} \right)
\; \left(\frac {\partial} {\partial u_i} \hat {\beta} \right)
' \]
\[ \hat {\beta} = J^{- 1} \; b \]
\[ u_i = w_i^2 \]
\[ J = [D' \; W \; D + S_D] = \sum_{i = 1}^ n u_i \; [d_i \; d_i'
+ S_i] \]
\[ b = D' \; W \; y = \sum_{i = 1}^ n u_i \; y_i \; d_i \]

\[ \frac {\partial} {\partial u_i} \hat {\beta} = \frac {\partial}
{\partial u_i} \left[J^{- 1} \; b \right] = J^{- 1} \; \left
\{\left[\frac {\partial} {\partial u_i} b \right] - \left[\frac
{\partial} {\partial u_i} J \right] \; J^{- 1} \; b \right\}
\]

\[ \frac {\partial} {\partial u_i} \hat {\beta} = J^{- 1} \; \left
\{\left[\frac {\partial} {\partial u_i} b \right] - \left[\frac
{\partial} {\partial u_i} J \right] \; \hat \beta \right\} \]

\[ \frac {\partial} {\partial u_i} \hat {\beta} = J^{- 1} \; \left
\{\left[\frac {\partial} {\partial u_i} \sum_{i = 1}^ n u_i
\; y_i \; d_i \right] - \left[\frac {\partial} {\partial u_i}
\sum_{i = 1}^ n u_i \; [d_i \; d_i' + S_i] \right] \; \hat \beta
\right\} \]

\[ \frac {\partial} {\partial u_i} \hat {\beta} = J^{- 1} \; \left
\{[y_i \; d_i] - [d_i \; d_i' + S_i] \; \hat \beta \right\} \]

\[ \ccv (\hat \beta) = \frac N {N - 1} \; J^{- 1} \; \sum_{i =
1}^ n u_i^2 \; \left\{[y_i \; d_i] - [d_i \; d_i' + S_i] \;
\hat \beta \right\} \; \left\{[y_i \; d_i] - [d_i \; d_i' +
S_i] \; \hat \beta \right\} ' \; J^{- 1} \]
} with the next sandwich estimator of covariance
\[ \ccv (\hat \beta) = \frac N {N - 1} \; J^{- 1} \left\{\sum_{i
= 1}^ n w_i^4 \; \left[e_i d_i - S_i \hat \beta \right] \left
[e_i d_i - S_i \hat \beta \right] ' \right\} J^{- 1} \]
where
\[ J = D' \; W \; D + S_D = \sum_{i = 1}^ n w_i^2 \; [d_i \; d_i'
+ S_i] \et e_i = y_i - d_i' \hat \beta . \]

The connections with ridge regression and with total least squares
are evident.

The derivation of above $ \ccv (\hat \beta) $ has been made by
the infinitesimal jackknife of Jaeckel (1972) according to Rey
(1983, Eq. 4.36) and sandwich estimators are surveyed by Freedman
(2006). Their lack of efficiency is notorious as investigated
by Kauermann and Carrol (2001) and it can be associated to the
high stochastic variability of the central factor between the
curled braces. In order to stabilize this factor, we imply independence
conditions similar to those assumed in ordinary linear regression.
We approximate the central factor and eventually obtain the covariance
estimator (see further details at Note \ref{not4})
\beq {eq16} \ccv (\hat \beta) = \frac N {N - p} \; J^{- 1} \left
\{\sum_{i = 1}^ n w_i^4 \; \left[s^2 d_i d_i' + (S_i \hat \beta)
(S_i \hat \beta) ' \right] \right\} J^{- 1} \eeq
with
\[ N = \frac {\left(\sum_{i = 1}^ n w_i^2 \right)^2} {\sum_{i
= 1}^ n w_i^4} \et s^2 = \frac {\sum_{i = 1}^ n w_i^2 \; (y_i
- d_i \hat \beta)^2} {\sum_{i = 1}^ n w_i^2} . \]

\subsection{Robust generalised least squares.}

The extension to the robust context is natural and follows closely
the above presentation for the one-dimension location estimator
by the set of equations \rf{eq10}.

Given previous estimates of $ (\beta, s) $, the weights and the
new $ (\beta, s) $ are estimated according to
\beq {eq17} \left\{
\begin {array} l \mbox {Given} \; \beta \; \mbox {and} \; s, \\
\mbox {} \\
w_i = \left(1 + u_i^4 \right)^ {- 1 / 4}, \ou \; u_i = \frac
{(y - D \; \beta)_i} {k_3 \; s}, \\
\mbox {} \\
J = \sum_{i = 1}^ n w_i^2 \; [d_i \; d_i' + S_i], \\
\mbox {} \\
{\beta} = J^{- 1} \; D' \; W \; y, \\
\mbox {} \\
s^2 = \left[{\sum_{i = 1}^ n w_i^2 \; (y_i - d_i \hat \beta)
^2} \right] / \left[{\sum_{i = 1}^ n w_i^2} \right] .
\end {array} \right.
\skp {abc \} .} \eeq

The associated covariance estimator
\[ \ccv (\hat \beta) = k_2^2 \; \frac N {N - p} \; J^{- 1} \left
\{\sum_{i = 1}^ n w_i^4 \; \left[s^2 d_i d_i' + (S_i \hat \beta)
(S_i \hat \beta) ' \right] \right\} J^{- 1} \ou N = \frac {\left
(\sum_{i = 1}^ n w_i^2 \right)^2} {\sum_{i = 1}^ n w_i^4} \]
is corrected for underestimation. It turns out that the constants
$k_3$ of and $k_2$ are the same as for the one-dimension location
estimator. Hence, the values of Table \ref{t2} are relevant in
regression as well.

\section{Singular Value Decomposition, SVD}

We deal with the minimisation \rf{eq2},
\beq {eq18} \armn {A \; B'} {\parallel X - A \; B' \parallel},
\eeq
where the approximation has rank $p$
\[ \upc X m n \upa {\approx} \upc A m p \upc {B'} p n \upa. \]
As we know, matrices $A$ and $B$ can be scaled and rotated in
any convenient way; to limit this indeterminacy, we impose the
weak condition that
\[ A \spc \mbox {be an orthonormal basis.} \]

\subsection{Total SVD, in view of statistical applications}

First, in order to illustrate the inconsistencies that we mentioned
in the introduction, we consider a simple numerical example. Let
the data matrix to be analysed by singular value decomposition
be next $X$. Except for a minor perturbation at level of the second
decimal place and an outlier, it is of {\sf Rank 1}; the last
entry should have been $x_{5, 3} = 15$.
\[ X = \left(
\begin {array} {ccc} 1 & 2 & 3 \\
2 & 4 & 6 \\
3 & 6 & 9 \\
4 & 8 & 12 \\
5 & 10 & 0 \\
\end {array} \right) + 0.001 \, \left(
\begin {array} {ccc} - 92 & 3 & - 17 \\
48 & 6 & - 8 \\
26 & - 4 & - 64 \\
8 & - 2 & 92 \\
17 & - 3 & 0 \\
\end {array} \right) = \left(
\begin {array} {ccc} 0.908 & 2.003 & 2.983 \\
2.048 & 4.006 & 5.992 \\
3.026 & 5.996 & 8.936 \\
4.008 & 7.998 & 12.09 \\
5.017 & 9.997 & 0 \\
\end {array} \right) . \]
Applying the ordinary (least squares) SVD, the approximation is
$A \; B' $
\[ \lgd {Ordinary SVD} {1, nonrobust} \spc X \spc \approx \spc
A \; B' = \left(
\begin {array} {ccc} 1.167 & 2.326 & 2.574 \\
2.364 & 4.710 & 5.212 \\
3.527 & 7.027 & 7.777 \\
4.741 & 9.445 & 10.45 \\
2.536 & 5.053 & 5.592 \\
\end {array} \right) . \]
The outlier has completely spoiled the evaluation.

Initialising with this least squares approximation, we bring the
attention on the two norms used in the evaluations of $A$ and
$B'$. Running \rhs{eq17} with $k_3 = 1.5$, they induce robust
weights on the data entries,
\[ \armn {B'} {\parallel X - A \; B' \parallel} \spc \rightarrow
\spc \mbox {weights} = \left(
\begin {array} {ccc} 0.931 & 0.984 & 0.999 \\
0.997 & 1.000 & 0.997 \\
0.999 & 0.988 & 0.977 \\
1.000 & 0.982 & 0.980 \\
0.039 & 0.009 & 0.010 \\
\end {array} \right) \]
and
\[ \armn A {\parallel X - A \; B' \parallel} \spc \rightarrow \spc
\mbox {weights} = \left(
\begin {array} {ccc} 0.987 & 0.975 & 0.869 \\
0.998 & 0.949 & 0.870 \\
0.997 & 0.953 & 0.868 \\
0.997 & 0.956 & 0.867 \\
0.997 & 0.955 & 0.867 \\
\end {array} \right) . \]
Clearly these two norms do not match. Moreover, this does not
help for better defining the norm of \rh{eq18}. As will soon be
seen and in order to avoid the above incompatibility between the
norms, we allocate the same set of weights for the evaluations
of
\[ \armn {B'} {\parallel X - A \; B' \parallel} \et \armn A {\parallel
X - A \; B' \parallel} . \]

The second issue we raised in the introduction is with respect
to the decomposition into an $ A \; B' $ product. Each of the
two components is estimated ``as if" the other one was properly
``known" rather than ``estimated".

We now further detail and, momentarily for the simplicity, we
place ourself in the ordinary least squares context. The squared
Frobenius norm of \rh{eq2} can be written as the double summation
\beq {eq19} \armn {\{A, B \}} {\sum_{i = 1}^m \sum_{j = 1}^n
(x_{i j} - a_i' \; b_j)^2} \eeq
where the residuals of all approximated $X$-entries are minimized.
Clearly, \rh{eq19} guarantees a good approximation, namely a good
estimation of the $ A \; B' $ product. However in very many cases
and, specifically for statistics in most low-rank reduction applications
of SVD, we are interested in good estimations of $A$ and $B$.
The qualities of the two components have the utmost importances
rather than the quality of their product. Then, it leads to minimize
\beq {eq20} \armn {\{A, B \}} {\sum_{i = 1}^ m \sum_{j = 1}^
n \left[(x_{i j} - a_i' \; b_j)^2 + a_i' \; \ccv (b_j) \; a_i
+ b_j' \; \ccv (a_i) \; b_j \right]} \eeq
that has a total least squares flavour; the minimisations takes
place on the $p$ columns of $A$ and the $p$ rows of $B'$. Accordingly,
we call ``Total SVD", the singular value decomposition based on
the norm \rh{eq20} (or \rh{eq21}, further on).

Seeing that the alternating process of \rh{eq3} and \rh{eq4} can
yield informations on $ \ccv (b_j) $ and $ \ccv (a_i) $, the present
set-up is less general although many of the arguments of Golub
and van Loan (1980), most of them also reviewed in the classroom
note of Nievergelt (1994), still hold. In some sense, \rh{eq20}
describes a trade-off between best approximating and best estimating
the components of the approximation. The increased complexity
of \rh{eq20} compared to \rh{eq19} induces practical difficulties
that are already reported in Gabriel and Zamir (1979). In fact,
the information we obtain on $ \ccv (b_j) $ and $ \ccv (a_i) $
is less rich than the full estimations of these matrices. For
instance, the estimation of $B$ by \rh{eq3} provides $ n$ covariance
matrices of dimensions $ \dm p p$, whereas $ \ccv (b_j) $ in \rh{eq20}
has dimensions $ \dm m m$. This mismatch of dimensions clearly
displays a difficulty; we estimate $A$ and $B$ column-wise, but
we use them row-wise in \rh{eq20}. However we can make use of
the variances of the $A$- and $B$-entries, their covariances are
not suitably available. This leads to substitute
\[ \vov (a_i) = \diag \{\sigma^2 (a_{1, i}), ..., \sigma^2 (a_{m,
i}) \} \spc \mbox {in place of} \spc \ccv (a_i), \]
and similarly for $ \ccv (b_j) $. Hence, \rh{eq20} takes the form
\beq {eq21} \armn {\{A, B \}} {\sum_{i = 1}^ m \sum_{j = 1}^
n \left[(x_{i j} - a_i' \; b_j)^2 + a_i' \; \vov (b_j) \; a_i
+ b_j' \; \vov (a_i) \; b_j \right]} . \eeq
Cancellating $ \vov (b_j) $ and $ \vov (a_i) $, \rh{eq21} collapses
into the ordinary problem statement given by \rh{eq2}.

\subsection{Estimation of Total SVD}

In view of the past expositions, first we introduce the general
algorithm in a few lines as being the solving of a fixed point
problem.

Given previous estimates of $ \left\{A, \vov (a_i)_{i = 1...
m}, \rule {0mm} {4mm} B, \vov (b_j)_{j = 1... n} \right\} $,
the weights and the new $ \left\{A, \vov (a_i)_{i = 1... m},
\rule {0mm} {4mm} B, \vov (b_j)_{j = 1... n} \right\} $ are estimated
according to
\beq {eq22} \left\{
\begin {array} l \mbox {Given} \; \left\{A, \vov (a_i)_{i = 1...
m}, \rule {0mm} {4mm} B, \vov (b_j)_{j = 1... n} \right\} . \\
\mbox {} \\
\mbox {Evaluate the weights} \; w_{i j} . \\
\mbox {} \\
\mbox {Given $A$ and $\vov (a_i)_{i = 1...m}$, estimate $B$ and
$ \vov (b_j)_{j = 1...n}$} . \\
\mbox {} \\
\mbox {Given $ B$ and $ \vov (b_j)_{j = 1...n}$, estimate $A$
and $\vov (a_i)_{i = 1...m}$.}
\end {array} \right.
\skp {abc \} .} \eeq

We review the four steps and this gradually leads us to the compact
algorithm stated by \rhs{eq26}.

\begin {itemize}

\item Given $ \left\{A, \vov (a_i)_{i = 1... m}, \rule {0mm}
{4mm} B, \vov (b_j)_{j = 1... n} \right\} $.

In fact, the algorithm does not do any use of any assumed $ \vov
(b_j)_{j = 1... n} $. The past estimation of $A$ is entered in
the next algorithmic step and, for the re-estimation of $A$, we
enter a $ \vov (b_j)_{j = 1... n} $ as found while $B$ was estimated.

Hence, a better formulation would have been ``Given $ \left\{A,
\vov (a_i)_{i = 1... m}, \rule {0mm} {4mm}, B \right\} $". Taking
into account the definition $ \vov = \diag \{\sigma^2 (a_{1, i}),
..., \sigma^2 (a_{m, i}) \}, $ we observe that we only need to
enter the variances of the $A$-entries. Eventually, the proper
formulation is ``Given $ \left\{A, \sigma^2 (a_{i, k})_{i = 1...
m, k = 1..p}, \rule {0mm} {4mm}, B, s \right\} $". The last parameter,
$s$, is incidental; it is entered to speed up the evaluation of
the weights.

\item Evaluate the weights $ \; w_{i j} $.

As indicated, we allocate the same set of weights for the evaluations
of the two norms
\[ \armn {B'} {\parallel X - A \; B' \parallel} \et \armn A {\parallel
X - A \; B' \parallel} . \]
They are derived from the $ m \times n$ approximation residuals
$ (X - A \; B')_{ij} $ and depends on their scatter $s$ in a way
that is familiar to the reader. Formally, this gives the following
algorithmic piece.

\beq {eq23} \left\{
\begin {array} l \mbox {Given} \; \left\{A, \rule {0mm} {4mm}
B, s \right\} . \\
\mbox {} \\
\mbox {Evaluate the residuals} \; f_k = (X - A \; B')_{i j}, \spc
k = 1, ..., m \, n . \\
\mbox {} \\
\mbox {Iterate on the expressions :} \\
\mbox {} \spc \left\{
\begin {array} l w_k = \left(1 + [\frrac {f_k} {(k_3 \, s)}]
^4 \right)^ {- 1 / 4}, \\
\mbox {} \\
N = \left[\sum_{k = 1}^ {m \, n} \; w_k \right]^2 / \left[\sum_{k
= 1}^ {m \, n} \; w_k^2 \right], \\
\mbox {} \\
\nu = \left(m + n - \frac {p + 1} 2 \right) \; p, \\
\mbox {} \\
s^2 = \frac N {N - \nu} \; \left[\sum_{k = 1}^ {m \, n} \; w_k^2
\; f_k^2 \right] / \left[\sum_{k = 1}^ {m \, n} \; w_k^2 \right]
.
\end {array} \right.
\skp {abc \} .} \mbox {} \\
\mbox {} \\
\mbox {Allocate the $w_k$ :} \spc w_{i j} = w_k.
\end {array} \right.
\skp {abc \} .} \eeq
The iterations converge linearly and should be accelerated.

\item Given $A$ and $ \vov (a_i)_{i = 1... m} $, estimate $ B$
and $ \vov (b_j)_{j = 1... n} $.

The generalised least squares context of Section \ref{lgls} here
is relevant and, without further delay, we present the corresponding
algorithmic piece that solves
\[ \armn {B'} {\parallel X - A \; B' \parallel} . \]

\beq {eq24} \left\{
\begin {array} l \mbox {Given} \; \left\{A, \sigma^2 (a_{i k})_{i
= 1... m, k = 1..p} \rule {0mm} {4mm} \right\}, \\
\mbox {} \\
\mbox {Successively, for each of the columns $x_j$ of $X$,} \\
\mbox {} \\
\mbox {} \spc S_i = \diag \left\{\sigma^2 (a_{i 1}), ..., \sigma^2
(a_{i p}) \right\} \sdm p p. \\
\mbox {} \\
\mbox {} \spc J = \sum_{i = 1}^ m w_{i j}^2 \; [a_i \; a_i'
+ S_i] \sdm p p. \\
\mbox {} \\
\mbox {} \spc W_j = \diag \left(w_{1j}, ..., w_{nj} \right) \sdm
m m. \\
\mbox {} \\
\mbox {} \spc (B')_j = J^{- 1} \; A' \; W_j \; x_j \sdm p 1,
\\
\mbox {} \\
\mbox {} \spc s_j^2 = \left[{\sum_{i = 1}^ m w_{i j}^2 \; (x_{i
j} - a_i \; (B')_j)^2} \right] / \left[{\sum_{i = 1}^ m w_{i
j}^2} \right] . \\
\mbox {} \\
\mbox {} \spc N_j = \frrac {\left(\sum_{i = 1}^ m w_{i j}^2
\right)^2} {\sum_{i = 1}^ m w_{i j}^4} \mbox {} . \\
\mbox {} \\
\mbox {} \spc {\small \ccv \left[(B')_j \right] = k_2^2 \, \frac
{N_j} {N_j - p} \, J^{- 1} \left\{\sum_{i = 1}^ m w_{i j}^4
\left(s_j^2 a_i a_i' + [S_i \, (B')_j] [S_i \, (B')_j] ' \right)
\right\} J^{- 1}} . \\
\mbox {} \\
\mbox {} \spc \sigma^2 (b_{j k}) = \left(\ccv \left[(B')_j \right]
\right)_{k k} .
\end {array} \right.
\skp {abc \} .} \eeq
The notation distinguishing rows from columns requiring some attention,
as well as for convenience, we noted the matrix dimensions between
square brackets.

\item Given $ B$ and $ \vov (b_j)_{j = 1... n} $, estimate $A$
and $ \vov (a_i)_{i = 1... m} $.

Except for matrix transpositions, we mimic very closely \rh{eq24}
and present the corresponding algorithmic piece. It solves
\[ \armn {A'} {\parallel X' - B \; A' \parallel} . \]
Furthermore, we complete by imposing the condition that $ A $
be an orthonormal basis.

\beq {eq25} \left\{
\begin {array} l \mbox {Given} \; \left\{B, \sigma^2 (b_{j k})_{j
= 1... n, k = 1..p} \rule {0mm} {4mm} \right\}, \\
\mbox {} \\
\mbox {Successively, for each of the columns $x_i$ of $X'$,} \\
\mbox {} \\
\mbox {} \spc S_j = \diag \left\{\sigma^2 (b_{j 1}), ..., \sigma^2
(b_{j p}) \right\} \sdm p p. \\
\mbox {} \\
\mbox {} \spc J = \sum_{j = 1}^ n w_{i j}^2 \; [b_j \; b_j'
+ S_j] \sdm p p. \\
\mbox {} \\
\mbox {} \spc W_i = \diag \left(w_{1i}, ..., w_{nj} \right) \sdm
n n. \\
\mbox {} \\
\mbox {} \spc (A')_i = J^{- 1} \; B' \; W_i \; x_i \sdm p 1,
\\
\mbox {} \\
\mbox {} \spc s_i^2 = \left[{\sum_{j = 1}^ n w_{i j}^2 \; (x_{i
j} - b_j \; (A')_i)^2} \right] / \left[{\sum_{j = 1}^ n w_{i
j}^2} \right] . \\
\mbox {} \\
\mbox {} \spc N_i = \frrac {\left(\sum_{j = 1}^ n w_{i j}^2
\right)^2} {\sum_{j = 1}^ n w_{i j}^4} \mbox {} . \\
\mbox {} \\
\mbox {} \spc {\small \ccv \left[(A')_i \right] = k_2^2 \, \frac
{N_i} {N_i - p} \, J^{- 1} \left\{\sum_{j = 1}^ n w_{i j}^4
\left(s_i^2 b_j b_j' + [S_j \, (A')_i] [S_j \, (A')_i] ' \right)
\right\} J^{- 1}} . \\
\mbox {} \\
\mbox {} \spc \sigma^2 (a_{i k}) = \left(\ccv \left[(A')_i \right]
\right)_{k k} ' \\
\mbox {} \\
\mbox {} \spc \mbox {Transform $A$ into an orthonormal basis.}
\end {array} \right.
\skp {abc \} .} \eeq
At the end of the convergence of \rhs{eq22}, the orthonormalisation
of $A$ becomes an identity operation.

\end {itemize}

Summing up, the above algorithmic parts obtain the Total SVD solution
as a fixed point of $ \left\{A, \sigma^2 (a_{i, k})_{i = 1...
m, k = 1..p}, \rule {0mm} {4mm}, B, s \right\} $ according to
\beq {eq26} \left\{
\begin {array} l \mbox {Evaluate the weights $ w_{i j}$ by \rhs
{eq23} .} \\
\mbox {} \\
\mbox {Estimate $ B$ and by \rhs {eq24} .} \\
\mbox {} \\
\mbox {Estimate $A$ by \rhs {eq25} .}
\end {array} \right.
\skp {abc \} .} \eeq
Before leaving this presentation, it must be noted that some difficulties
of convergence can occur (see Note \ref{not5}).

We apply the variants of the above algorithm on the $ \dm 5 3$
example and specially draw the attention on the value taken by
$x_{5, 3} $ (that should be $x_{5, 3} = 15 $). We already met
with the usual SVD result,
\[ \lgd {Ordinary SVD} {1, nonrobust} \spc X \spc \approx \spc
A \; B' = \left(
\begin {array} {ccc} 1.167 & 2.326 & 2.574 \\
2.364 & 4.710 & 5.212 \\
3.527 & 7.027 & 7.777 \\
4.741 & 9.445 & 10.45 \\
2.536 & 5.053 & 5.592 \\
\end {array} \right) . \]
Taking into account the limited precision of product components
$A$ and $B$ adds little to the quality of the resulting approximation,
\[ \lgd {Total SVD} {1, nonrobust} \spc X \spc \approx \spc A \;
B' = \left(
\begin {array} {ccc} 1.078 & 2.147 & 2.376 \\
2.183 & 4.349 & 4.813 \\
3.257 & 6.489 & 7.180 \\
4.377 & 8.721 & 9.651 \\
2.342 & 4.666 & 5.164 \\
\end {array} \right) . \]
However, becoming robust induces a major quality improvement,
\[ \lgd {Ordinary SVD} {1, robust : $k_3=1$} \spc X \spc \approx
\spc A \; B' = \left(
\begin {array} {ccc} 1.005 & 2.000 & 2.983 \\
2.018 & 4.016 & 5.989 \\
3.012 & 5.995 & 8.941 \\
4.018 & 7.997 & 11.93 \\
5.021 & 9.995 & 14.91 \\
\end {array} \right), \]
and this is even better when Total SVD is estimated,
\[ \lgd {total SVD} {1, robust : $k_3=1$} \spc X \spc \approx \spc
A \; B' = \left(
\begin {array} {ccc} 0.9990 & 1.989 & 2.987 \\
2.009 & 3.999 & 6.006 \\
2.998 & 5.969 & 8.963 \\
4.034 & 8.032 & 12.06 \\
5.020 & 9.995 & 15.01 \\
\end {array} \right) . \]

Thinking at biplot applications, we analyse the European Health
and Fertility data treated at Section 5.1 of Croux et al (1993).
It is their $ \dm {16} 9$-Table 1 set and clearly it has columns
of very different scalings,
\[ \mbox {European Data} \spc = \spc \left(
\begin {array} {rrrrr} - 0.1 & 48 & ... & 3440 & 6 \\
1.8 & 50 & ... & 2716 & 7 \\
0.2 & 47 & ... & 3593 & 6 \\
... & ... & ... & ... & ... \\
1.9 & 49 & ... & 3218 & 8 \\
0.5 & 51 & ... & 3499 & 7 \\
0.2 & 48 & ... & 3421 & 5 \\
\end {array} \right) . \]
The scalings are so different that it makes little sense to act
as if some homoscedasticity could be assumed. Hence, we transform
the data set and the columns become centered and with variances
1,
\[ \mbox {European Data} \spc = \spc \left(
\begin {array} {rrrrr} - 0.8778 & 0.26812 & ... & 0.43654 & -
0.0768 \\
2.2616 & 1.4938 & ... & - 2.2925 & 0.53751 \\
- 0.3821 & - 0.3447 & ... & 1.0132 & - 0.0768 \\
... & ... & ... & ... & ... \\
2.4268 & 0.88097 & ... & - 0.4003 & 1.1518 \\
0.11359 & 2.1067 & ... & 0.65893 & 0.53751 \\
- 0.3821 & 0.26812 & ... & 0.36492 & - 0.6911 \\
\end {array} \right) . \]
The treatment allocate a weight to each of the entries,
\[ \lgd {Total SVD} {2, robust : $k_3=1$} \spc \mbox {Weights}
= \left(
\begin {array} {ccccc} 0.95895 & 0.77783 & ... & 0.94071 & 0.98998
\\
0.27558 & 0.32983 & ... & 0.20418 & 1.00000 \\
0.79620 & 0.97667 & ... & 0.56217 & 1.00000 \\
... & ... & ... & ... & ... \\
0.98847 & 0.99972 & ... & 0.99998 & 0.99144 \\
1.00000 & 0.21158 & ... & 0.65800 & 0.94428 \\
0.99977 & 0.96668 & ... & 0.93092 & 0.93200 \\
\end {array} \right), \]
and we do not observe any very small weight. We are entitled to
say that no entry is clearly outlying. The 5 lowest weights are
\[ 
\begin {array} {| ccl |}
\hline
\mbox {Weight} & \mbox {Entry} & \mbox {Country, Factor} \\
\hline
0.18181 & (7, 9) & \mbox {H, baby underw.} \\
0.18522 & (2, 7) & \mbox {AL, inhab. doc.} \\
0.20418 & (2, 8) & \mbox {AL, calorie} \\
0.20607 & (13, 3) & \mbox {SU, women \%} \\
0.21158 & (15, 2) & \mbox {YU, give birth} \\
\hline
\end {array} \]
Contrary to the observation of Croux et al (1993), we do not have
to eliminate by down-weighting the full fourteenth row corresponding
to Turkey; what properly fits with the Rank 2 decomposition, properly
contributes to our estimation.

Of course, the situation becomes more extreme when we increase
the level of robustness. Now we clearly find outlying entries,
\[ \lgd {Total SVD} {2, robust : $k_3=0.5$} \spc \mbox {Weights}
= \left(
\begin {array} {ccccc} 0.16691 & 0.18247 & ... & 0.27318 & 0.25611
\\
0.04368 & 0.06851 & ... & 0.04143 & 0.60169 \\
0.26271 & 0.79319 & ... & 0.12728 & 0.99098 \\
... & ... & ... & ... & ... \\
0.07989 & 0.99997 & ... & 0.99992 & 0.50327 \\
0.99110 & 0.04645 & ... & 0.14016 & 0.30331 \\
0.99951 & 0.16474 & ... & 0.38905 & 0.42217 \\
\end {array} \right) \]
and the above 5 low weights decrease,
\[ 
\begin {array} {| ccl |}
\hline
\mbox {Weight} & \mbox {Entry} & \mbox {Country, Factor} \\
\hline
0.03469 & (7, 9) & \mbox {H, baby underw.} \\
0.03422 & (2, 7) & \mbox {AL, inhab. doc.} \\
0.04143 & (2, 8) & \mbox {AL, calorie} \\
0.03915 & (13, 3) & \mbox {SU, women \%} \\
0.04645 & (15, 2) & \mbox {YU, give birth} \\
\hline
\end {array} \]
As is ordinary in the field of robustness, the analyst must master
the tools he uses. A complementary feature is how vary the eigenvalues
of $ \Lambda = \diag \{\lambda_1, ..., \lambda_p \} $ in the ordinary
SVD of the approximation $A \; B'$; this is reported at Table
\ref{t3}.

\btb {Eigenvalues of $A \; B'$ for the European Health and Fertility
data.} {t3} {|c|c|cc|ccc|}
\hline
& {Rank 1} & \multicolumn 2 {c |} {{Rank 2}} & \multicolumn 3
{c |} {{Rank 3}} \\
\hline
& & & & & & \\
{Ordinary SVD} & 8.5194 & 8.5194 & 5.7931 & 8.5194 & 5.7931 &
3.2940 \\
[6pt] {Total SVD} & & & & & & \\
$k_3 = \infty$ & 7.5963 & 7.7213 & 4.7238 & 7.1071 & 3.6495 &
3.1878 \\
$k_3 = 2.0$ & 7.4691 & 7.5167 & 4.5743 & 7.1802 & 3.6663 & 3.1619
\\
$k_3 = 1.0$ & 7.3074 & 7.3593 & 4.0388 & 7.0150 & 3.3955 & 3.0195
\\
$k_3 = 0.5$ & 7.2501 & 6.4301 & 3.1360 & 6.3143 & 3.098 & 1.0491
\\
\hline
\etb

\newpage

\mbox{} \\
{\bf Numerical notes}

\begin{enumerate}

\item \label{not1} {\bf The Gaussian numerical sample.}

For a sample of size $ m$, we represent the Gaussian distribution
by its $ m$ quantiles defined as follows: we generate the points
\[ x_i \; \sth {\; \int_{- \infty}^ {x_i} \phi (x) \; d \, x =
\frac {i - 1 / 2} m}, \ou \phi (x) = \exp^{- x^2 / 2} / \sqrt
{2 \; \pi} . \]

\item \label{not2} {\bf Breakdown point $ \bp_a $ by \rh{eq13}.}

We add $k$ points to a sample of the standard Gaussian, constructed
as above by $ m$ quantiles. The $k$ new points are located far
away, namely at $x = 10^6$. Then, $ \mbox {\sf Offset}_n $ varies
continuously with $k$ and we can evaluate $ \bp_a$ for any $a$
value. For numerical convenience, we take $a = \sigma_x$ and evaluate
$ \bp_1 $.

Clearly, the estimated $ \bp_a$ depends on the selected $a$-value.
However, this dependence is very moderate. The curve $ \bp_a (a)
$ is linear for small $a$ and, fairly suddenly, displays a sharp
increase that defines a quasi-asymptote.

\item \label{not3} {\bf Asymptotic estimations.} Consider an estimator
$t = t (m) $ that depends on the sample size $ m$ used to represent
the standard Gaussian distribution. We assume the model
\[ t (m) = t (\infty) + \frac {t_1} {m + t_2} \]
and expect the parameter $t_2$ to be small.

The three sample sizes $ m = 100, 300$ and $900$ are used in the
experiment. A model fit delivers the asymptotic value $t (\infty)
$.

In the linear regression context of \rhs{eq17}, rather than implementing
with respect to the two factors of the product $k_3 = k_1 \times
k_2$, it is convenient to parametrize on $k_3$ and approximate
$k_2$ by an expression such as
\[ \ln k_2 \approx ({0.4762 - 0.8465 \; \ln k_3 + 0.4554 \; \ln^2
k_3}) / ({1 - 0.3425 \; \ln k_3}) . \]

\item \label{not4} {\bf Covariance estimator, \rh{eq16}.}

In addition to the approximation
\[ \sum_{i = 1}^ m w_i^4 \; \left[e_i d_i - S_i \hat \beta \right]
\left[e_i d_i - S_i \hat \beta \right] ' \approx \sum_{i = 1}
^m w_i^4 \; \left[s^2 d_i d_i' + (S_i \hat \beta) (S_i \hat \beta)
' \right] \]
where $ s^2 = \overline {e_i^2}, $ we modified the scaling factor
obtained by the infinitesimal jackknife, $ N / (N - 1) $, into
the $ N / (N - p) $ familiar in ordinary linear regression. The
above approximation and the latter modification have been supported
by simulation runs with arbitrary weights and diagonal matrices
$S_i$. The covariance estimator \rh{eq16} tends to slightly overestimate
the true covariance when the right term of \mbox{$ \left[s^2
d_i d_i' + (S_i \hat \beta) (S_i \hat \beta) ' \right] $} dominates
the left one, on the average.

\item \label{not5} {\bf Computation of Total SVD by \rhs{eq26}.}

The experimental estimation of the fixed point of \rhs{eq26} has
presented hazards. It sometimes occurs that the mapping contracts
in a fairly small vicinity of the solution $ \left\{A, \sigma^2
(a_{i, k})_{i = 1... m, k = 1..p}, \rule {0mm} {4mm}, B, s \right
\} .$ Eventually, we have been solving
\[ \armn {\{A, B \}} {\sum_{i = 1}^ m \sum_{j = 1}^ n \left[(x_{i
j} - a_i' \; b_j)^2 + t \; \left(a_i' \; \vov (b_j) \; a_i +
b_j' \; \vov (a_i) \; b_j \right) \right]} \]
for a parameter $t$ varied from $t = 0$ to $t = 1$. We apply a
predictor-corrector method of continuation built around the fixed
point algorithm.

\end{enumerate}

\[ \]

\mbox{} \\
{\bf References}

\mbox{} \\
Andrews, D.F., Bickel, P.J., Hampel, F.R., Huber, P.J., Rogers,
W.H. and Tukey, J.W. (1972) Robust Estimates of Location: Survey
and Advances, Princeton University Press, Princeton.

\mbox{} \\
Bradu, D. and Gabriel, K.R. (1978) The biplot as a diagnostic
tool for model of two-way tables, Technometrics, 20, 47-68.

\mbox{} \\
Coleman, D., Holland, P., Kaden, N., Klema, V. and Peters, S.C.
(1980) A system of subroutines for iteratively reweighted least
squares computations, ACM Transactions on Mathematical Software,
6, 327-336.

\mbox{} \\
Croux C., Filzmoser P., Pison G. and Rousseeuw P.J. (2003) Fitting
multiplicative models by robust alternating regressions, Statistics
and Computing, 13, 23-36.

\mbox{} \\
Freedman, D.A. (2006) On the so-called ``Huber sandwich estimator"
and ``Robust standard errors", The American Statistician, 60,
299-302.

\mbox{} \\
Gabriel, K.R. (1998) Generalised bilinear regression, Biometrika,
85, 689-700.

\mbox{} \\
Gabriel K.R. and Zamir S. (1979) Lower rank approximation of matrices
by least squares with any choice of weights, Technometrics, 21,
489-498.

\mbox{} \\
Golub, G.E. and van Loan, C.F. (1980) An analysis of the total
least squares problem, SIAM Journal Numerical Analysis, 17, 883-893.

\mbox{} \\
Gower, J.C. (2004) The geometry of biplot scaling, Biometrika,
91, 705-714.

\mbox{} \\
Hampel, F.R. (1968) Contributions to the theory of robust estimation,
Ph. D. Dissertation, University of California, Berkeley.

\mbox{} \\
Huber, P.J. (1964) Robust estimation of a location parameter,
Annals of Mathematical Statistics, 35, 73-101.

\mbox{} \\
Huber, P.J. (1972) Robust statistics: A review, Annals of Mathematical
Statistics, 43, 1041-1067.

\mbox{} \\
Huber, P.J. (1973) Robust regression: asymptotics, conjectures
and Monte Carlo, Annals of Statistics, 1, 799-821.

\mbox{} \\
Huber, P.J. (1996) Robust Statistical Procedures, Second Edition,
SIAM Monograph, Philadelphia, 68.

\mbox{} \\
Jaeckel, L.A. (1972) The infinitesimal jackknife, Bell Telephone
Laboratories, Memorandum MH-1215.

\mbox{} \\
Kauermann, G. and Carrol, R.J. (2001) A note on the efficiency
of sandwich covariance matrix estimation, Journal of the American
Statistical Association, 96, 1387-1396.

\mbox{} \\
Ke, Q. and Kanade, T. (2005) Robust L1 norm factorization in the
presence of outliers and missing data by alternative convex programming,
IEEE Conference on Computer Vision and Pattern Recognition (CVPR
2005), 739-746.

\mbox{} \\
Klema, V. (1978) Rosepack - Robust estimation package, ACM Signum
Letter, 13, 18-19.

\mbox{} \\
Le Roux, N.J. and Gardner, S. (2005) Analysing your multivariate
data as a pictorial: A case for applying biplot methodology? International
Statistical Institute Review, 73, 365-387.

\mbox{} \\
Nievergelt, Y. (1994) Total least squares: State-of-the-art regression
in numerical analysis, SIAM Review, 36, 258-264.

\mbox{} \\
Rey, W.J.J. (1978) Robust Statistical Methods, Lecture Notes in
Mathematics, 690, Springer-Verlag.

\mbox{} \\
Rey, W.J.J. (1983) Introduction to Robust and Quasi-robust Statistical
Methods, Universitext Series, Springer-Verlag.

\mbox{} \\
Tukey, J.W. (1960) A survey of sampling from contaminated distributions,
in Olkin, I. (Edt.), Contributions to Probability and Statistics,
Stanford University Press, 448-485.

\end {document}